\documentclass[pdflatex,sn-mathphys-num]{sn-jnl}

\usepackage{graphicx}%
\usepackage{multirow}%
\usepackage{amsmath,amssymb,amsfonts}%
\usepackage{amsthm}%
\usepackage{mathrsfs}%
\usepackage[title]{appendix}%
\usepackage{xcolor}%
\usepackage{textcomp}%
\usepackage{manyfoot}%
\usepackage{booktabs}%
\usepackage{algorithm}%
\usepackage{algorithmicx}%
\usepackage{algpseudocode}%
\usepackage{listings}%

\usepackage{caption}
\usepackage{subcaption}
\usepackage{booktabs}
\usepackage[table]{xcolor}
\usepackage{tabularx}

\theoremstyle{thmstyleone}%
\theoremstyle{thmstyletwo}%

\theoremstyle{thmstylethree}%

\begin{document}

\title[]{MRSeqStudio: MRI Sequence Design and Simulation \textit{as a Service} in a Free and Open-Source Web Platform}


\author{
\small
\makebox[\textwidth][c]{%
\textbf{Pablo Villacorta-Aylagas}$^{1,*}$,
\textbf{Manuel Rodríguez-Cayetano}$^{1,2}$} \\[0.3em]
\makebox[\textwidth][c]{%
\textbf{Carlos Castillo-Passi}$^{3,4,5,6}$,
\textbf{Pablo Irarrazaval}$^{3,4,5}$} \\[0.3em]
\makebox[\textwidth][c]{%
\textbf{Federico Simmross-Wattenberg}$^{1,2}$,
\textbf{Carlos Alberola-López}$^{1,2}$} \\[0.8em]

\makebox[\textwidth][c]{%
$^{1}$Laboratorio de Procesado de Imagen, Universidad de Valladolid, Valladolid, Spain} \\
\makebox[\textwidth][c]{%
$^{2}$IBIOVall, Institute for Health Research, Valladolid, Spain} \\
\makebox[\textwidth][c]{%
$^{3}$Institute for Biological and Medical Engineering, Pontificia Universidad Católica de Chile, Santiago, Chile} \\
\makebox[\textwidth][c]{%
$^{4}$Millennium Institute for Intelligent Healthcare Engineering (iHEALTH), Santiago, Chile} \\
\makebox[\textwidth][c]{%
$^{5}$School of Biomedical Engineering and Imaging Sciences, King’s College, London, UK} \\
\makebox[\textwidth][c]{%
$^{6}$Department of Radiology, Stanford University, California, USA}\\[0.5em]

\small *Corresponding author: \texttt{pablo.villacorta@uva.es}
}


\abstract{MRI sequence prototyping increasingly relies on graphical design environments and numerical simulators to accelerate development and validation. While several platforms support interactive sequence construction, fully web-based solutions that combine integrated phantom management, high-fidelity Bloch simulation, and scalable multi-user deployment remain limited.

We present MRSeqStudio, a web-based platform for interactive MR sequence design and simulation. The tool adopts a block-based representation model with real-time visualization and native JSON/Pulseq export. Simulations are performed using the GPU-enabled Bloch simulator KomaMRI, which enables accurate modeling of arbitrary pulse sequences and phantoms within an installation-free architecture. The system separates front-end interaction from back-end simulation services to support concurrent multi-user access.

Sequence validity was assessed by comparing GRE and bSSFP implementations against equivalent sequences designed in mtrk and gammaSTAR. The resulting images showed minimal absolute differences and high mean structural similarity indices (SSIM).
Stress testing under burst-request conditions demonstrated stable performance with up to 100 concurrent users on a high-performance desktop deployment.
A comparative workflow analysis with mtrk and gammaSTAR further examined differences in representation models, parameter propagation strategies, and integration levels across platforms, highlighting the relative strengths and limitations of each tool.

Results indicate that MRSeqStudio provides a reliable and accessible environment for MR sequence prototyping, combining web-native deployment with Bloch-level simulation fidelity and integrated phantom visualization.}




\maketitle

\clearpage

\section{Introduction}\label{sec:intro}

Pulse sequence design plays a central role in the development and/or optimization of magnetic resonance imaging (MRI) protocols, with the aim of reducing scan time, improving image contrast or minimizing artifacts.
Originally, sequence design relied on vendor-specific frameworks, which were only available to scanner manufacturers~\cite{weine2023cmrseq,artiges2026mtrk} or to associated research institutions. In response to this, open-source initiatives for sequence design~\cite{magland2016,nielsen2018,weine2023cmrseq,cencini2024pulserver,konstandin2025gammastar,artiges2026mtrk} have emerged in recent years. 
As for sequence exchange,  Pulseq~\cite{layton2017pulseq,ravi2019pypulseq} has become the \emph{de facto} standard in the field. 

Complementarily, MRI simulation provides the ground for testing sequences without the need to access an actual MRI system; it also enables the analysis of specific acquisition phenomena 
and the synthesis of images for further usage, such as training machine learning models or generating signal dictionaries~\cite{castillo2023komamri}. 
In addition to these research-oriented uses, MRI simulation has also been widely adopted for educational~\cite{torheim1994,hacklander2005,tonnes2023virtmri} and training purposes~\cite{treceno2019,xanthis2019coreMRI}.

Taken together, these considerations highlight the importance of coordination between sequence design and simulation. When seamlessly connected, both processes enable a streamlined and efficient workflow from sequence conception to validation and testing. 

From a practical perspective, existing tools relying primarily on programming interfaces may present a significant entry barrier. To alleviate this, several platforms~\cite{magland2016,liu2017mrilab,xanthis2019coreMRI,castillo2023komamri,konstandin2025gammastar,artiges2026mtrk,tonnes2023virtmri} have already incorporated graphical user interfaces (GUIs), thereby lowering the technical threshold for adoption. 
Nevertheless, most of these tools focus predominantly on either sequence generation or simulation, and the integration of both functionalities within a unified graphical environment remains to be fully addressed. 

Notably, recent platforms have begun to bridge this gap by integrating both areas within web-based graphical environments, namely, gammaSTAR~\cite{konstandin2025gammastar} and the cloudMR environment which contains mrtk~\cite{artiges2026mtrk,montin2025cloudMR}. These solutions are intended for exploratory research, education, and early-stage validation, and represent key advances toward more integrated MRI workflows. However, in this context of sequence prototyping, additional requirements emerge. In particular, streamlined use without local installation, fast or near-real-time simulations enabling interactive feedback, and integrated digital phantom selection and visualization
remain areas with room for refinement. 

In response to these needs, we present MRSeqStudio, a web-based framework designed specifically for rapid pulse sequence prototyping. 
It integrates sequence design and simulation within a unified interface, providing real-time feedback, scanner parameter configuration, and interactive visualization of the sequence, the digital phantom (hereafter, the phantom), the selected slice, and the simulation results. 
Users can define global variables that propagate across all dependent sequence blocks, enabling high-level parameters to be set once and applied consistently throughout the sequence.
The tool extends the concept of sequence design ``as a service'' introduced in~\cite{artiges2026mtrk}, since it requires no local installation and offers an end-to-end workflow in which sequences are designed in the browser and simulated on a server.

Beyond the presentation of the platform itself, this work also includes a comparative evaluation of sequence design workflows across MRSeqStudio, gammaSTAR~\cite{konstandin2025gammastar} and mrtk~\cite{artiges2026mtrk}.
This comparison aims to provide an objective analysis of their respective sequence structures, interaction models, and integration strategies. Such an assessment may assist users in selecting the most appropriate tool depending on their specific research, educational, or prototyping needs.

The remainder of this paper is structured as follows. 
Section~\ref{sec:background} reviews related work and existing tools for sequence design and MRI simulation, with particular emphasis on the two platforms we use for comparison~\cite{konstandin2025gammastar,artiges2026mtrk}.
Section~\ref{sec:methods} describes the implementation of MRSeqStudio (\ref{sec:implementation}), the experiments used for validation and performance evaluation (\ref{sec:experiments}), and the protocol followed to conduct the comparative workflow analysis (\ref{sec:comparison}).
Section~\ref{sec:results} presents the results of the validation and scalability experiments (\ref{sec:results-experiments}), as well as a structured comparison of workflows across platforms (\ref{sec:results-comparison}).
Section~\ref{sec:discussion} discusses the implications of these findings.
Finally, Section~\ref{sec:conclusion} summarizes the main contributions and outlines future directions.

\section{Background and Related Work}\label{sec:background}

Several graphical tools for sequence design have been proposed over the years. One of the earliest widely adopted frameworks was SequenceTree~\cite{magland2016}, which represents pulse sequences as hierarchical trees of modular building blocks. A similar tree-based abstraction is followed by the graphical editor of JEMRIS~\cite{stocker2010jemris} and more recently by gammaSTAR~\cite{konstandin2025gammastar}. In contrast, mtrk~\cite{artiges2026mtrk} allows users to directly manipulate the temporal diagram of the sequence, providing a more waveform-oriented interaction model.

Regarding MRI simulators, tools can be broadly classified into three categories: 
(1) sequence-specific analytical simulators~\cite{treceno2019, tonnes2023virtmri,weine2024cmrsim} evaluate the expressions of the most common sequences, providing speed but limited modeling of complex phenomena; 
(2) Bloch simulators~\cite{stocker2010jemris,xanthis2014mrisimul,liu2017mrilab,castillo2023komamri,loktyushin2021mrzero,weine2024cmrsim} solve the Bloch equations for each isochromat (i.e., spin~\cite{castillo2023komamri}), achieving high-fidelity simulations at the expense of computational cost, and thus drawing on GPU acceleration in several implementations~\cite{xanthis2014mrisimul,liu2017mrilab,xanthis2019coreMRI,castillo2023komamri,loktyushin2021mrzero,weine2024cmrsim,nurdinova2025gpu};
(3) EPG-based simulators~\cite{loktyushin2021mrzero,endres2024mrzero_epg} use the extended phase graph formalism~\cite{weigel2015epg} to provide an analytical description of the Bloch equations in the Fourier domain, offering accuracy and computational efficiency, but reduced flexibility for arbitrary phantoms or complex RF pulses. 

Currently, two platforms stand out for integrating MRI sequence design and simulation within a web-oriented ecosystem: mtrk and gammaSTAR.

mtrk is embedded within the cloudMR framework~\cite{montin2025cloudMR}, which interconnects independent modules to reproduce complete MRI experiments, including sequence design~\cite{artiges2026mtrk}, coil configuration~\cite{giannakopoulos2026marie}, simulation~\cite{montin2020camrie}, reconstruction~\cite{uecker2015bart} and SNR evaluation~\cite{montin2019mroptimum}. Within this environment, mtrk provides a GUI for sequence construction. The fundamental unit in mtrk is the ``event'' (e.g., gradient, RF pulse, or ADC), which is explicitly placed on a conventional sequence time diagram.
Its JSON-based Sequence Description Language (SDL) is human-readable, allowing users to develop sequences by (1) writing the SDL file through its Python API, (2) using the graphical interface, (3) generating the file from any JSON-compatible language, or (4) manually editing the SDL file. 
In addition to supporting export to Pulseq (which enables execution on real MRI systems), the tool relies on a so-called ``driver sequence''~\cite{artiges2026mtrk} to enable direct execution on vendor hardware. This driver sequence interprets the SDL file and translates it into native real-time scanner events.
The cloudMR simulation module, CAMRIE~\cite{montin2020camrie}, is a web-based application that uses a modified version of KomaMRI, which is publicly available\footnote{KomaNYU.jl: \url{https://github.com/cloudmrhub/KomaNYU.jl}.}.

gammaSTAR~\cite{konstandin2025gammastar} is a fully web-based application for interactive MRI sequence design. Its core abstraction is the ``blueprint'', an entity composed of parameters that can represent low-level elements (e.g., waveforms, time samples), high-level protocol parameters (e.g., TE, TR, flip angle), or intermediate derived quantities. Blueprints are hierarchically organized in a tree-like structure , where relationships between elements are explicitly defined and parameter dependencies can be propagated across levels.
Like mtrk, gammaSTAR supports export to Pulseq, as well as a sequence driver for direct interaction with real scanner environments through a dedicated execution interface that translates the internal sequence representation into vendor-specific instructions. 
gammaSTAR is natively connected to the MRzero simulator~\cite{loktyushin2021mrzero}, allowing sequences designed in the front-end  to be executed directly without intermediate conversion steps. Its architecture emphasizes modularity and parameter propagation through predefined relationships between blueprints, facilitating protocol-driven configuration and reuse of composite sequence elements.

Despite their significant advances, several practical aspects remain partially addressed. 
Although full web-based deployment would not represent a fundamental technical limitation, neither platform currently provides a completely installation-free solution that grant access to all functionalities. 
Furthermore, beyond simulation speed itself, overall interactivity depends on efficient communication between front-end and back-end, which may affect real-time prototyping workflows. 
Integrated phantom management is also limited: neither platform currently offers a dedicated interface for selecting and visualizing phantoms, including the ability to handle dynamic (moving) phantoms. Section~\ref{sec:results-comparison}  discusses in more detail these issues.

\section{Methods}\label{sec:methods}

\subsection{The Application}\label{sec:implementation}

We outline the most relevant technical details of the application design and implementation. The complete source code is publicly available in a dedicated repository\footnote{\url{https://github.com/pvillacorta/MRSeqStudio}.}.

\subsubsection{Design Requirements}
MRSeqStudio must support rapid pulse sequence prototyping while maintaining high physical accuracy and usability. The main requirements are:
\begin{itemize}
    \item \textbf{Block-based sequence design.} Users should be able to construct sequences from elementary blocks which should be combinable, rearrangeable, and groupable into composite units to build sequences of arbitrary complexity.
    \item \textbf{User-defined global variables.} Users should be able to define variables that can be referenced and manipulated throughout the sequence, enabling consistent parametrization and high-level control over sequence parameters.
    \item \textbf{Interactive visualization and feedback.} Users should be able to inspect the sequence through a time diagram and a 3D slice viewer. They should be able to switch between immediate slice visualization and full volume rendering using the integrated simulation back-end.
    \item \textbf{Web-based interface.} The platform should allow users to design and simulate sequences entirely in the browser, without the need for local installation.
    \item \textbf{Integrated simulation back-end.} The system should provide real-time or near-real-time simulations on the server to give immediate feedback on sequence behavior.
    \item \textbf{Phantom selection and scanner parameter configuration:} Users should be able to select phantoms, adjust scanner parameters (e.g., $B_0$, $G_{max}$, $S_{max}$), and visualize their impact on simulated sequences.
    \item \textbf{Scalability and multi-user support.} The platform should be capable of handling multiple concurrent users while maintaining performance.
\end{itemize}

\subsubsection{Architectural Design and Selected Technologies} \label{sec:architectural}

\begin{figure}[ht]
    \begin{centering}
        \captionsetup{, margin=0.3cm}
        \includegraphics[width=\textwidth]{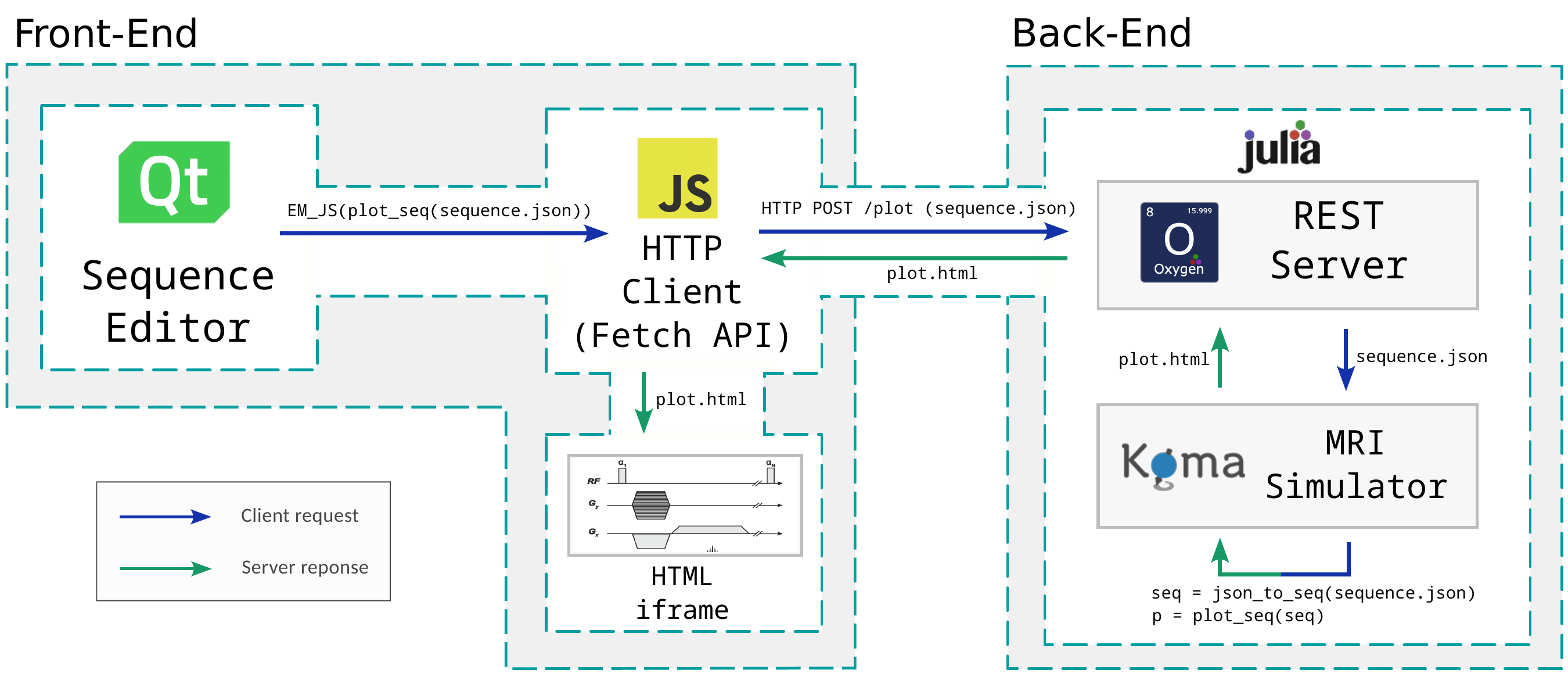}
        \caption[]{Schema of a client request for a sequence plot. A user action in the front end triggers an HTTP request to the RESTful API; the server generates the corresponding visualization and returns HTML content, which is embedded in the viewer panel. Blue arrows denote requests and green arrows denote responses.}
        \label{fig:architecture}
    \end{centering}
\end{figure}

The application follows a client–server architecture (Fig.~\ref{fig:architecture}), where the browser communicates with the server through HTTP requests to a RESTful~\cite{richardson2007restful} API that provides access to back-end resources. 

Fig.~\ref{fig:architecture} also illustrates the division of the application into different modules, deployed either on the front-end or on the back-end. 
The former is composed of four components:
(1) the sequence editor itself, developed in Qt/QML, chosen for its suitability in building cross-platform GUIs, including deployment to WebAssembly~\cite{webassembly};
(2) a sequence diagram viewer; 
(3) a phantom viewer; and 
(4) a simulation results viewer. 
These three viewers are generated on the back-end and embedded in the front-end via HTML iframes. In addition, the phantom viewer provides a second visualization mode based on orthogonal slices, implemented directly in the front end using vtk.js.

The four front-end modules are coordinated via JavaScript, which communicates with the server through HTTP requests issued using the \texttt{fetch} API to the RESTful interface. Actions performed in the sequence editor trigger requests for either sequence plotting or simulation. The server responds with HTML content that is embedded into the corresponding viewer panel. The phantom viewer is initialized at startup and updated when a new phantom is selected, following the same request–response mechanism.

Regarding the back-end design, a Bloch-based simulator was selected, as this class of methods provides the most general framework for modeling arbitrary pulse sequences, gradient waveforms, and phantom configurations (see Section~\ref{sec:background}).
Specifically, KomaMRI was chosen due to its high computational performance compared to alternative Bloch-based frameworks~\cite{castillo2023komamri,endres2024mrzero_epg,nurdinova2025gpu,pablo2026mrm}.
Since KomaMRI is implemented in Julia~\cite{bezanson2017julia}, the remainder of the back-end has also been developed using this programming language for smooth integration. Specifically, we rely on the following packages: 
KomaMRI.jl for MRI simulations and for generating sequence and phantom visualizations; 
MySQL.jl for database access and user management; and 
Oxygen.jl for the implementation of the RESTful API. 
In addition, several auxiliary libraries are employed, including JSON3.jl for JSON serialization and deserialization; CUDA.jl (or the corresponding package depending on the server hardware) for GPU acceleration; DBInterface.jl for database abstraction; SwaggerMarkdown.jl for API documentation; and JWTs.jl for authentication and token management.

\subsubsection{Sequence Representation and Interaction Model} \label{sec:sequence}

\textbf{Block Abstraction} \\ 
In the proposed framework, a ``block'' is defined as a set of events that occur simultaneously within a given time interval. These events may include RF pulses, gradients, and data acquisition windows (ADC). The block abstraction thus represents the smallest unit of a sequence in which concurrent physical processes are specified, providing a slightly higher level of abstraction than the events in mtrk or the blueprints in gammaSTAR (see Section~\ref{sec:background}).

Block types are distinguished by the combination of events they contain, and each type is assigned a numerical code (Table~\ref{tab:blocks}). Code 0 corresponds to groups, a special type of block whose only parameters are the number of repetitions and the indices of its child blocks, and which enable the definition of loops (see Section~\ref{sec:resolution}).

\begin{table}[ht]
\centering
\rowcolors{2}{gray!10}{white}
\begin{tabular}{c c l}
\toprule
\rowcolor{gray!20}
\textbf{Code} & \textbf{Block Name} & \textbf{Contained Events} \\
\midrule
0 & Group & Repetition container + child blocks \\
1 & \texttt{Ex} & RF + optional gradients \\
2 & \texttt{Delay} & No events (time interval) \\
3 & \texttt{Dephase} & Gradients only \\
4 & \texttt{Readout} & Gradients + ADC \\
5 & \texttt{EPI\_ACQ} & Cartesian EPI readout (auto gradients + ADC) \\
6 & \texttt{GRE} & Complete gradient-echo module \\
\bottomrule
\end{tabular}
\caption{Sequence block types and associated events.}
\label{tab:blocks}
\end{table}

\begin{figure}[ht]
    \centering
    \begin{subfigure}[b]{0\textwidth}
        \phantomcaption
        \label{fig:listmodel}
    \end{subfigure}
    \begin{subfigure}[b]{0\textwidth}
        \phantomcaption
        \label{fig:json}
    \end{subfigure}
    \includegraphics[width=\textwidth]{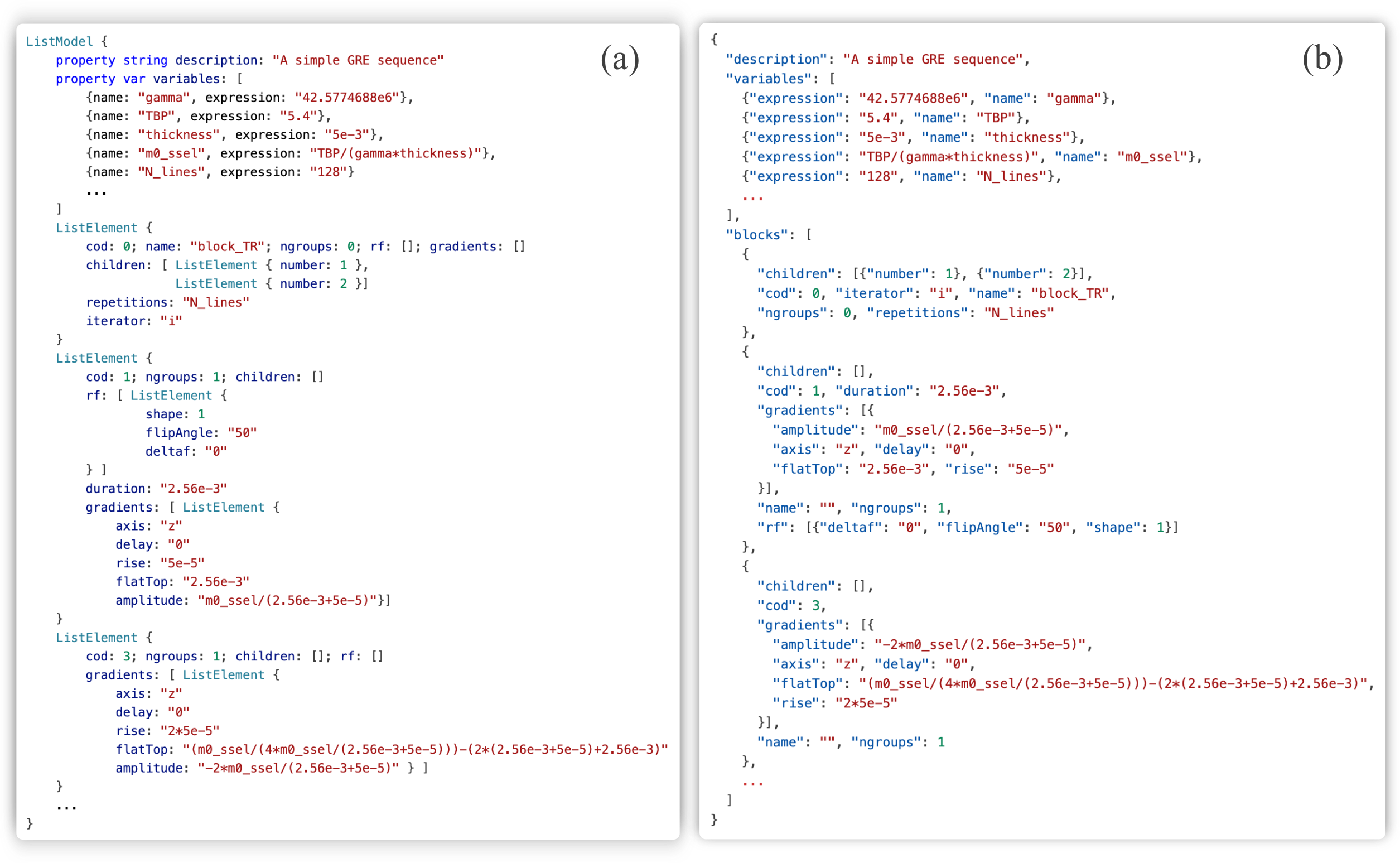}
    \caption{Example of internal sequence representation in (a) QML \texttt{ListModel} and (b) its exportable JSON format, which is also the format understood by the back-end and suitable for sharing sequences. The figure is illustrative; some fields present in the full data structures have been omitted for clarity.}
    \label{fig:representation}
\end{figure}

\textbf{Internal Sequence Representation} \\
The sequence is internally represented in QML using a \texttt{ListModel} (Fig.~\ref{fig:listmodel}), where each element corresponds to a single block. This model is linear.  
All blocks share a common set of properties, regardless of their specific type (except for groups). Some parameters are not required for certain block classes, but they are defined uniformly and initialized to zero when unused.

In addition to the block list, the sequence data structure maintains a collection of user-defined global variables. These variables can be defined through mathematical expressions, which may themselves reference other variables. Any configurable field of a block can be then specified using an expression involving these variables (see the example of Fig~\ref{fig:listmodel}, where the definition of the variable \texttt{m0\_ssel} involves other three user-defined variables and is then used to define the strength of the slice-selection gradient).

The \texttt{ListModel} can be directly exported to JSON (Fig.~\ref{fig:json}), which preserves the complete high-level structure of the sequence, including block hierarchy, loops, and symbolic expressions. JSON therefore constitutes the native exchange format of the application and the interface with the back-end simulator.

Export to Pulseq is also supported. In this case, the sequence is first serialized to JSON in the front-end and sent to the server, where it is parsed and converted into a \texttt{KomaMRI} sequence object. The corresponding Pulseq file is then generated using KomaMRI’s writer and returned to the client; this workflow follows the same scheme depicted in Fig.~\ref{fig:architecture}. Unlike the JSON representation, the Pulseq format encodes the fully resolved low-level event sequence and does not retain high-level constructs such as blocks, groups, or variables.

\textbf{Interaction Model} \\
The internal sequence representation can be manipulated through the Qt/QML GUI. Blocks are displayed as draggable elements arranged along a temporal axis.
Each block exposes a configuration panel where its parameters can be edited. 
Changes performed in the interface immediately update the underlying \texttt{ListModel}.
Global variables are managed through a dedicated panel in the front-end, where users can define, modify, or remove variables, and any update automatically propagates to all dependent variables and block parameters through expression re-evaluation.

\subsubsection{Sequence Resolution and Simulation}\label{sec:resolution}

Alongside the sequence JSON file, a separate JSON object which contains scanner parameters (e.g., maximum gradient amplitude and slew rate) is sent to the back-end, which enforces compliance with these hardware constraints. 
During resolution, all block parameters are evaluated, including those expressed symbolically via global variables or referencing the current loop iteration, and loops/groups are expanded to generate a fully instantiated low-level sequence.
Before simulation, the hierarchical block structure is resolved into a KomaMRI sequence object, composed of low-level events and with all expressions evaluated and loops expanded. 
The sequence is then executed on the server, producing either a sequence diagram, or a full Bloch simulation.

\subsubsection{Application Graphical Interface Overview}
\begin{figure}[ht]
    \centering
    \includegraphics[width=\textwidth]{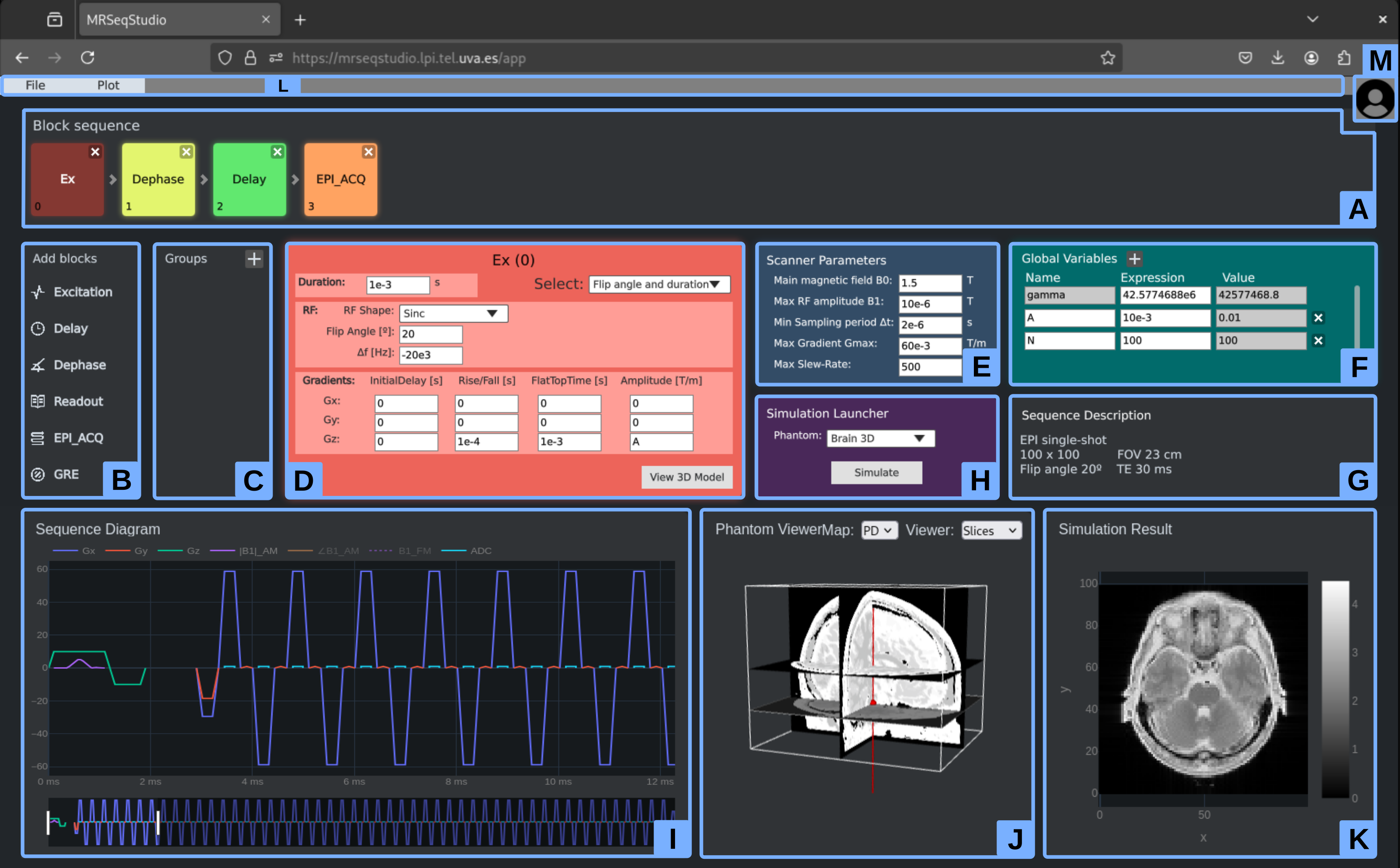}
    \caption{Application main layout, divided into panels (blue rectangles) for  sequence composition (A-C), block configuration (D),  scanner parameter and variable definition (E, F), phantom selection and simulation control (H), and  visualization (I-K). The example shows the design and simulation of a GE-EPI sequence on a 3D brain phantom.}
    \label{fig:interface} 
\end{figure}

Fig.~\ref{fig:interface} shows the main interface, which is organized into multiple panels. 
Panel A contains the blocks comprising the sequence. 
Panel B allows the user to add different types of blocks (see Table~\ref{tab:blocks}). 
Panel C supports group creation and replication. 
Panel D displays the configuration menu for a specified block. 
Panel E is used to define the scanner parameters. 
Panel F provides the environment where global variables can be defined as either numerical values or mathematical expressions. 
Panel G allows the user to define a textual sequence description. 
Panel H enables phantom selection and the execution of MRI simulations in the back-end. Simulations are performed over the selected phantom and using the sequence designed in the GUI.
Panels I-K correspond to the modules (2-4) mentioned in Section~\ref{sec:architectural}, and display, respectively, the sequence temporal diagram, the phantom, and the MRI simulation output.
Panel L provides file handling for loading and saving sequence and scanner files.
Button M opens the user panel, which allows managing the current session, logging out, and accessing stored sequences, results, and account information.

Three illustrative experiments, provided as Supplementary Videos S1–S3, demonstrate sequence loading, modification of protocol-related parameters, and visualization of simulation outputs within the MRSeqStudio interface.
In particular, Video S3 showcases motion-related simulation capabilities supported by the KomaMRI back-end~\cite{pablo2026mrm} and their integration into the application workflow.

\subsection{Experiments}\label{sec:experiments}
We describe the experiments performed for validation (Section \ref{sec:validation}) and multi-user scalability (Section \ref{sec:scalability}). The source code of these experiments is available in another repository\footnote{\url{https://github.com/pvillacorta/MRSeqStudioExperiments}.}.

\subsubsection{Simulation Output Validation}\label{sec:validation}
For validation, the output of MRSeqStudio was compared with those obtained with the two established platforms~\cite{artiges2026mtrk,konstandin2025gammastar} described in Section~\ref{sec:background}. Two representative sequences (GRE and bSSFP) were considered. Both were initially defined in gammaSTAR: the bSSFP sequence corresponds to a built-in preset, whereas the GRE sequence was constructed following the platform’s tutorial entitled “How to Build a Simple Sequence”\footnote{\url{https://gammastar.mevis.fraunhofer.de/\#/help}.}. This same GRE tutorial was also used in Section~\ref{sec:comparison} as the basis for the comparative workflow assessment, focusing on the procedural steps and user interactions required to construct the sequence in each tool. The sequences were subsequently replicated in the other platforms.
Specifically, these steps were followed:
(1) each sequence was generated within the three platforms, yielding three tool-specific JSON representations;
(2) the sequences were exported to Pulseq using the respective Pulseq writers provided by each tool;
(3) the resulting Pulseq files were executed in the same simulator (KomaMRI) under identical phantom (3D brain) and scanner conditions;
(4) absolute difference and structural similarity (SSIM~\cite{zhou2004ssim}) maps were obtained for image comparison.

\subsubsection{Performance and Multi-User Scalability}\label{sec:scalability}

To assess the scalability of MRSeqStudio under multi-user scenarios, Apache JMeter~\cite{jmeter} was used to generate HTTP requests that emulate concurrent users interacting with the application. The elapsed time from request submission to server response was measured. Preset sequences provided by the application were used for these tests.

Each simulated user executed a loop repeated 10 times, consisting of the following sequence of requests: (1) Login, (2) get preset sequence list, (3) get the full description of a specific sequence (in this case, \texttt{GE\_EPI\_singleshot}), (4) plot the sequence, and (5) simulate the sequence. 
For this last simulation request, the measured response times correspond to the initial sequence submission and the server response in the case that the submission is accepted (either started or added to the queue), but not the duration of the simulation itself since it entirely depends on the back-end simulator.

The experiment was performed with with 5, 10, 50, and 100 simultaneous users, all issuing their requests concurrently.

The computer in which the application (both front-end and back-end) was deployed for tests had an AMD Ryzen 9 9950X3D 16-Core Processor and 64 GB of RAM.

\subsection{Comparative Workflow Assessment}\label{sec:comparison}

To evaluate the usability and workflow of the three platforms, a spoiled GRE sequence was constructed in each tool, documenting the steps, required interactions, challenges encountered, and workflow characteristics. 

For gammaSTAR, the sequence was built by following its tutorial ``How to Build a Simple Sequence'' referred to above; the steps included: (1) starting from an empty template and adding slice-selective excitation and line readout blueprints; (2) defining the kernel duration and adapting start times to satisfy protocol constraints; (3) introducing loop structures to iterate over the required number of phase encoding steps, and parameterizing line-specific variables accordingly.

For mtrk, sequence creation followed a mixed approach that combines strategies (2) and (4) described in Section~\ref{sec:background}: the sequence was initially designed using the GUI, exported to JSON for manual adjustments, and then re-imported into the GUI to verify and refine the design. Specifically, the process consisted of: (1) adding the events (RF pulses, gradients, and ADCs) into the diagram to construct a single TR; (2) assigning numerical values or expressions to the event parameters to match the intended protocol; (3) grouping all events into a TR block and repeating as needed to build the full sequence.

For MRSeqStudio, the sequence was assembled in the GUI as follows: (1) placing the blocks that constitute a single TR (excitation-dephase-delay-readout-delay-dephase); (2) defining global variables to characterize high-level sequence parameters (e.g., TE, TR, flip angle) and propagate them to lower-level block attributes such as RF amplitudes and timing fields (e.g., \texttt{m0\_ssel} in Fig.~\ref{fig:representation}); (3) grouping all blocks into a loop structure to define the number of repetitions corresponding to the desired number of phase encoding lines.

In this case, we focused on documenting the workflow itself (how sequences are created, parameterized, and repeated) rather than on the exact simulation outcomes.

\section{Results}\label{sec:results}
We report the validation and performance results, along with a comparison of sequence design workflows across mtrk, gammaSTAR, and MRSeqStudio.

\subsection{Experiments}\label{sec:results-experiments}
\subsubsection{Simulation Output Validation}
\begin{figure}[ht]
    \centering
    \includegraphics[width=\textwidth]{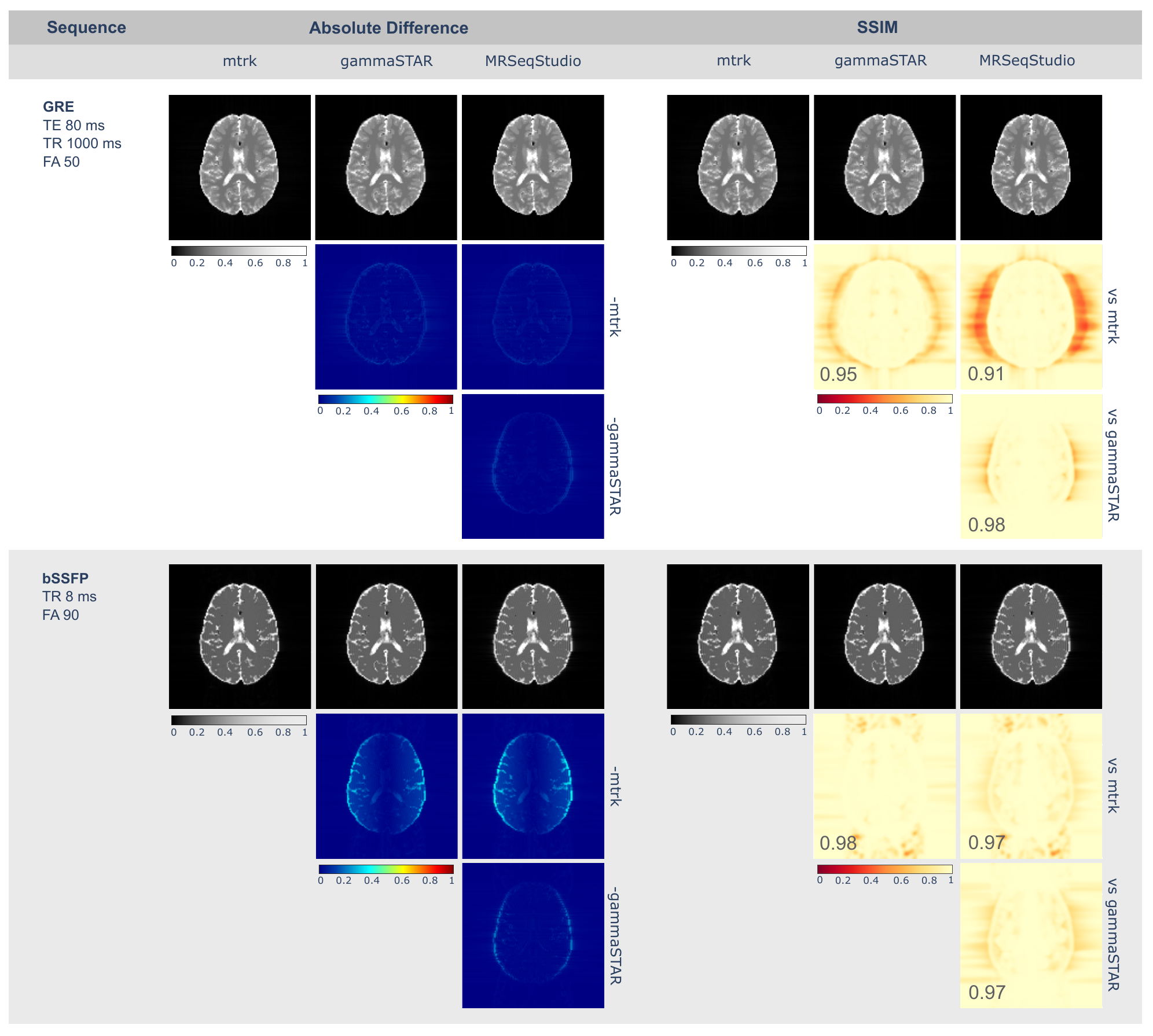}
    \caption{Validation of the GRE and bSSFP sequences generated with mtrk, gammaSTAR, and MRSeqStudio. Magnitude images are shown alongside absolute difference and SSIM maps for pairwise comparisons between tools.}
    \label{fig:validation} 
\end{figure}

Fig.~\ref{fig:validation}, which is divided into four panels, shows the magnitude images obtained from the GRE and bSSFP sequences generated by each platform, along with absolute difference and SSIM maps for pairwise comparisons. The left panels display absolute difference maps, while the right panels show SSIM maps with the corresponding mean SSIM value for each comparison. In each panel, the first row presents the magnitude images from each tool, and the subsequent rows illustrate comparisons between the tool in the column and the tool in the row (e.g., the element in row 2, column 2 corresponds to the comparison between gammaSTAR and mtrk).
Magnitude images, absolute difference maps, and SSIM maps are all represented on a 0–1 scale. For SSIM maps, values close to 0 indicate low similarity, whereas values close to 1 indicate high similarity.
All images were initially rectangular (128 × 256 pixels) and were subsequently cropped to 128 × 128 pixels, centered on the anatomical region of interest.

The results demonstrate strong agreement between sequences generated by mtrk, gammaSTAR, and MRSeqStudio, with minimal absolute differences and mean SSIM values consistently above 0.9. Most discrepancies are confined to background regions outside the brain, where low signal levels amplify the relative impact of small numerical variations. Within anatomically relevant regions, differences remain negligible.

\subsubsection{Performance and Multi-User Scalability}

\begin{figure}[ht]
    \centering
    \includegraphics[width=\textwidth]{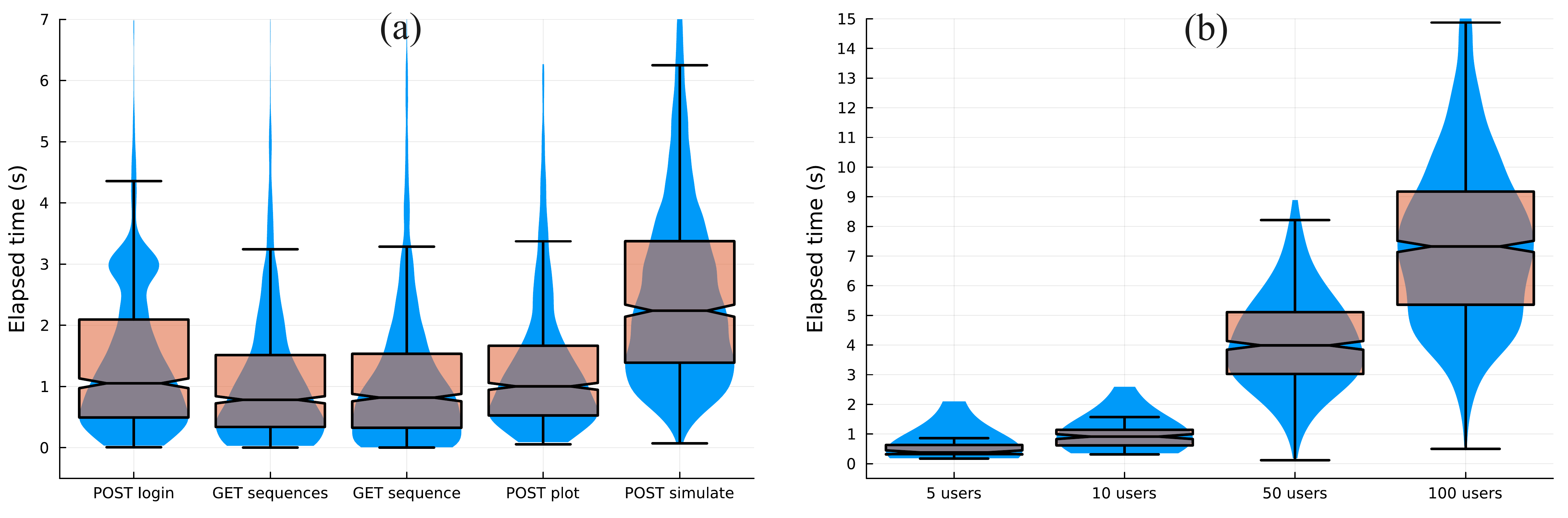}
    \begin{subfigure}[b]{0\textwidth}
        \phantomcaption
        \label{fig:scalability-a}
    \end{subfigure}
    \begin{subfigure}[b]{0\textwidth}
        \phantomcaption
        \label{fig:scalability-b}
    \end{subfigure}
    \caption{Multi-user scalability evaluation. (a) Response time distributions for each request category under 100 simultaneous users (10 iterations per user). (b) Total time required to complete the five-request interaction sequence for 5, 10, 50, and 100 concurrent users. All experiments were conducted under synchronized start conditions.}
    \label{fig:scalability} 
\end{figure}

Figure~\ref{fig:scalability} summarizes the response times measured under concurrent multi-user scenarios. 
Figure~\ref{fig:scalability-a} presents the distribution of response times for the five request categories with 100 simultaneous users. For lightweight operations such as login and sequence retrieval, average response times remained around 1.1–1.4~s, with limited dispersion. The plot request exhibited a comparable mean response time of 1.32~s.
As expected, the simulation submission request showed the highest latency (mean 2.59 s), reflecting server-side validation and job scheduling.

Figure~\ref{fig:scalability-b} shows the total time required to complete the five-request interaction sequence as a function of the number of simultaneous users. Although dispersion increases with concurrency, the median total interaction time remains controlled across all configurations.

\subsection{Comparative Workflow Assessment} \label{sec:results-comparison}
Table~\ref{tab:comparison}, derived both from Section~\ref{sec:background} and from observations made during the comparative workflow analysis (Section~\ref{sec:comparison}), summarizes the features of each platform, including core sequence design elements such as representation models and parameterization strategies, as well as complementary aspects that influence the overall user workflow, such as visualization and integration capabilities.
The table is structured with individual features listed as rows, and the three platforms as columns. 
Features in which one platform offers a comparatively more complete or integrated implementation are highlighted in bold to facilitate the identification of distinguishing characteristics.

Overall, all three platforms provide real-time sequence visualization and support native JSON/Pulseq export. They differ primarily in sequence representation (event diagram, tree-based, or block-based structure), looping and repetition mechanisms, and the degree of integration with phantom management and scanner configuration. Some features, such as arbitrary waveform definition, remain integrated only in mtrk and gammaSTAR, with MRSeqStudio not yet supporting this functionality.

For mtrk, multiple features are described as being integrated within the cloudMR environment. This reflects the fact that mtrk operates as part of a broader modular ecosystem in which functionalities such as phantom management, scanner configuration, and constraint validation are handled at the environment level. At the time of writing, the fully integrated cloudMR deployment is not yet publicly accessible as a unified platform, with release expected in the near future. Consequently, the table refers to these features at the ecosystem level without further implementation-specific detail.

These observations provide an objective overview of workflow-related characteristics and platform capabilities, which will be further interpreted in Section~\ref{sec:discussion}.

\begin{table}[ht]
\centering
\footnotesize
\def\arraystretch{1.5}%
\begin{tabularx}{\textwidth}{X X X X}
\toprule
\rowcolor{gray!20}
\textbf{Feature} & \textbf{mtrk}  & \textbf{gammaSTAR} & \textbf{MRSeqStudio} \\
\midrule

\rowcolor{gray!10}
\textbf{Sequence representation model} &
Event diagram with grouped blocks &
Tree-based blueprint structure &
Block-based grouped structure \\

\textbf{Sequence visualization} &
Real-time interactive plot &
Real-time interactive plot &
Real-time interactive plot \\

\rowcolor{gray!10}
\textbf{Looping/repetition definition} &
Manual repetition via TR block &
Explicit loop blueprints &
Manual repetition via TR block \\

\textbf{Preset sequences and modules} &
Basic preset sequences (GRE, SE) &
\textbf{Extensive preset sequences and reusable blueprint models} &
Preset sequences (GRE, bSSFP, SE), and reusable aquisition blocks (EPI) \\

\rowcolor{gray!10}
\textbf{Phantom selection} &
Integrated within the cloudMR environment &
Available from the MRzero simulation back-end (not yet from the gammaSTAR GUI) &
\textbf{Fully integrated web-based phantom selection}\\

\textbf{Phantom visualization} &
Integrated within the cloudMR environment &
No dedicated phantom visualization interface &
\textbf{Integrated interactive 3D/4D anatomical viewer (slice preview capabilities)} \\

\rowcolor{gray!10}
\textbf{Simulation execution model} &
Integrated within the cloudMR environment &
Web front-end with locally deployed MRzero simulation back-end &
\textbf{Fully integrated web-based execution} \\

\textbf{Arbitrary waveform definition} &
\textbf{Integrated} &
\textbf{Integrated} &
Not yet integrated \\

\rowcolor{gray!10}
\textbf{Sequence export formats} &
Native JSON-based \& Pulseq &
Native JSON-based \& Pulseq &
Native JSON-based \& Pulseq \\

\textbf{Communication with MRI system} &
\textbf{Via Pulseq and sequence driver} &
\textbf{Via Pulseq and sequence driver} &
Via Pulseq\\

\rowcolor{gray!10}
\textbf{Protocol parameter propagation} &
User-defined global variables and explicit parameter expressions &
\textbf{Protocol-driven parameters pre-defined within composite blueprints} &
User-defined global variables and explicit parameter expressions  \\

\textbf{Definition of scanner parameters} &
Integrated within the cloudMR environment &
Configurable within the dedicated \texttt{sys} blueprint &
Configurable within the \textit{Scanner Parameters} panel \\

\rowcolor{gray!10}
\textbf{Scanner constraints validation} &
Integrated within the cloudMR environment &
Integrated directly within the front-end &
Integrated within KomaMRI back-end \\

\textbf{Required actions for simple sequence construction} &
(1) Event placement, (2) amplitude and timing-related manual configuration, and (3) grouping within TR block  &
\textbf{(1) Blueprint placement, (2) timing-related configuration, and (3) looping blueprint addition} &
(1) Block placement, (2) amplitude and timing-related manual configuration, and (3) grouping within TR block \\

\bottomrule
\end{tabularx}

\normalsize 

\caption{Comparison of structural, functional, and interaction-related features of mtrk, gammaSTAR, and MRSeqStudio.}
\label{tab:comparison}
\end{table}

\clearpage

\section{Discussion}\label{sec:discussion}
For the validation experiment, the three platforms show strong agreement, and small image differences cannot be attributed to differences in simulation accuracy, as all sequences were executed using the same simulator (KomaMRI) under identical conditions. Rather, they stem from subtle variations in how each sequence was designed within the respective software environments. Although all implementations are functionally correct, differences in sequence construction approaches (see Section~\ref{sec:results-comparison}) make it difficult to obtain identical internal representations across platforms, reflecting different but equally valid approaches.

In terms of application performance under multi-user conditions in MRSeqStudio, the experiment represents a worst-case scenario in which all users initiate their requests simultaneously, causing burst arrivals at each request stage. Despite this unfavorable condition, MRSeqStudio maintains reasonable response times even at 100 concurrent users, compatible with uninterrupted user interaction according to classical usability thresholds~\cite{nielsen1994usability}.
Minor variations in latency are observed for the login operation (see Fig.~\ref{fig:scalability-a}), likely related to database access operations.
Note that the experiments were conducted on a desktop workstation which, albeit powerful, cannot match the performance of a dedicated server infrastructure. 
Nevertheless, additional performance optimization is possible. The current architecture relies on a single back-end process responsible for handling API requests and coordinating with the database and the MRI simulation processes. A more scalable deployment strategy, such as distributing requests across multiple processes and/or threads\footnote{\url{https://oxygenframework.github.io/Oxygen.jl/stable/\#Multithreading-and-Parallelism}.} or introducing more advanced resource management mechanisms, could further reduce latency under heavier workloads.

The comparative workflow assessment reveals both shared features, such as real-time sequence visualization and Pulseq export, and important differences in sequence modularity, parameter propagation, and workflow integration, which influence design efficiency and potential application scenarios.

All three platforms support Pulseq export, which provides a standardized format sufficient for sequence deployment on MRI systems and simulators. In addition, mtrk and gammaSTAR include vendor-specific sequence drivers, which offers direct control for end-to-end execution while introducing additional complexity.

Beyond execution, the platforms differ in usability and simulation capabilities. mtrk offers an intuitive event diagram interface, allowing users to manipulate sequence events directly, which may reduce the learning curve for new users. MRSeqStudio, in contrast, stands out for its fully integrated phantom selection, interactive 4D phantom viewer, and near real-time simulation feedback, supporting rapid prototyping and verification of sequences. As of today, it is the only platform that provides a full web-based interface with multiuser concurrency and no need for local installation. This is the reason why the experiment described in Section~\ref{sec:scalability} could not be carried out in the other two platforms. 

gammaSTAR distinguishes itself through its extensive library of preset sequences and reusable modules, facilitating rapid sequence construction and reducing manual configuration. Both mtrk and MRSeqStudio provide fewer pre-configured modules, which increases the effort required during sequence design. Expanding the library of preset modules in these platforms could enhance workflow efficiency and consistency. Additionally, gammaSTAR benefits from automatic propagation of protocol parameters through its sequence blueprints, reducing the number of manual adjustments needed; similar functionality in mtrk and MRSeqStudio could be achieved by leveraging additional pre-configured modules.

Rather than being mutually exclusive, the three platforms can be seen as complementary. For instance, front-ends such as gammaSTAR or MRSeqStudio could be paired with back-end simulators like KomaMRI or MRzero to create hybrid prototyping and simulation environments. Although these integrations are not yet fully realized, they illustrate the potential for a global workflow ecosystem in which tools interoperate seamlessly.

\section{Conclusion}\label{sec:conclusion}
This work presented MRSeqStudio, a web-based platform for MRI sequence design and simulation based on a block-oriented representation model. The application was validated through practical use and tested in multi-user environments, demonstrating its suitability for collaborative and interactive prototyping workflows.

A functional comparison with mtrk and gammaSTAR, two established web-oriented platforms integrating sequence design and simulation, highlighted both shared foundations (such as Pulseq interoperability and real-time visualization) and key differences in representation models, modularity, execution strategies, and integration capabilities. Within this assessment, MRSeqStudio stood out for its integrated phantom selection, interactive 4D visualization, near real-time simulation feedback and full web-based access with user concurrency,  while also revealing areas for further development,
including support for complex sequences and arbitrary waveforms\footnote{See status in: \url{https://github.com/pvillacorta/MRSeqStudio/issues/6}.}, pre-defined intermediate global variables (e.g., gradient moments , dephasing angles, or optimal durations), improved documentation, and clinical validation through testing Pulseq-exported sequences on real MRI scanners. Addressing these points would enhance the platform's versatility and usability for advanced sequence design.

Overall, the results indicate that these platforms are best understood as complementary rather than interchangeable tools. Depending on user needs and workflow priorities, different solutions may be preferable. Continued convergence toward standardized formats and improved cross-platform integration may enable a more unified ecosystem for MRI sequence prototyping and deployment.

\clearpage

\section*{Supplementary information}

Additional supporting material may be found online in the Supporting Information tab for this article.

\begin{itemize}
    \item \textbf{Video S1}: EPI sequence simulation experiment. A GRE-EPI sequence is loaded into the application GUI and some of its parameters are modified. Then, simulation is launched and the result is plotted within the ``Results'' panel.
    \item \textbf{Video S2}: Spin Echo simulation experiment. A SE sequence is loaded into the GUI and three consecutive simulations are conducted, with varying TE and TR values.
    \item \textbf{Video S3}: Time-of-flight experiment. Two different sequences (a GRE-EPI and a bSSFP) are consecutively loaded and simulated over a cylindrical phantom with flow inside.
\end{itemize}

\section*{Declarations}

\subsection*{Funding}

This work was supported by the Agencia Estatal de Investigaci\'on with grants PID2020-115339RB-I00, TED2021-130090B-I00 and CPP2021-008880, and to Fundaci\'on La Caixa with grant HR22-00533. P.V.A. has been funded under the 2024 UVa predoctoral contracts call, co‑funded by Banco Santander.” Additional funds correspond to ANID-Chile grants FONDECYT 1210747, FONDECYT Post-doctoral 3240576 and Millennium Science Initiative Program - ICN2021\_004.

\subsection*{Conflict of interest/Competing interests}
The authors declare that they have no competing interests.

\subsection*{Data and Materials availability}
The source code of the application has been deposited in a public GitHub repository: \\
\url{https://github.com/pvillacorta/MRSeqStudio}
. This repository includes instructions for deploying the application locally.

The source code of the experiments has been deposited in another public GitHub repository: 
\url{https://github.com/pvillacorta/MRSeqStudioExperiments}.

The MRSeqStudio platform is also accessible online at: \\
\url{https://mrseqstudio.lpi.tel.uva.es}, where users can access the application by registering an account and logging in through a standard web browser.

\subsection*{Authors' contributions}
P.V.A. was responsible for the conceptualization, implementation, and software development of the tool, as well as for performing the validation and workflow assessment experiments and writing the first draft of the manuscript. M.R.C. and F.S.W. contributed to the conceptualization and software implementation, and carried out the multi-user scalability experiments, also participating in reviewing and refining the manuscript. C.C.P. and P.I. assisted in revising the manuscript and clarifying specific aspects related to KomaMRI. C.A.L. supervised the work throughout all its stages and contributed substantially to shaping and improving the final version of the manuscript.
\\\\
All authors have read and approved the final manuscript.

\subsection*{Acknowledgments}
The authors would like to sincerely thank Anaïs Artiges for willingness to discuss the design and implementation of mtrk, as well as for her support in facilitating its comparison with MRSeqStudio. The authors also thank Eros Montin for his guidance regarding the cloudMR ecosystem, and Simon Konstandin for his clarifications concerning gammaSTAR and his interest in the development of MRSeqStudio.

The authors acknowledge the financial support of the Agencia Estatal de Investigación, Fundación La Caixa, Banco Santander, and ANID-Chile through the projects mentioned in the Funding section. The authors also wish to thank the members of the ISMRM Iberian Chapter for testing the application and providing valuable feedback.

\clearpage


\end{document}